\documentstyle[pra,aps,preprint,graphics,epsfig]{revtex}

\begin{document}

\tightenlines

\draft

\preprint{Version: ANU, \today}

\title{Quantum field effects in coupled atomic and molecular
Bose-Einstein condensates}

\author{J.~J. Hope}

\address{Department of Physics and Theoretical Physics, Australian 
National University,\\
ACT 0200, Australia.}

\maketitle

\begin{abstract}

This paper examines the parameter regimes in which coupled atomic and
molecular Bose-Einstein condensates do not obey the Gross-Pitaevskii
equation.  Stochastic field equations for coupled atomic and molecular
condensates are derived using the functional positive-P
representation.  These equations describe the full quantum state of
the coupled condensates and include the commonly used Gross-Pitaevskii
equation as the noiseless limit.  The model includes all interactions
between the particles, background gas losses, two-body losses and the
numerical simulations are performed in three dimensions.  It is found
that it is possible to differentiate the quantum and semiclassical
behaviour when the particle density is sufficiently low and the
coupling is sufficiently strong.
\end{abstract}

\section{Introduction}

One of the surprising discoveries that has been made since the
experimental production of a Bose-Einstein condensate (BEC) in a
weakly interacting gas~\cite{Anderson95} has been the fact that
virtually all of their properties can be described using the
Gross-Pitaevskii equation (GPE).  This semiclassical approximation
ignores any dynamics in the quantum statistics of the field.  There
has been much interest recently in the production of a molecular
Bose-Einstein condensate (MBEC) from the photoassociation of an atomic
BEC of a weakly interacting dilute gas
\cite{Levi00,Drummond98,Heinzen00,Wynar00,Julienne98,Javanainen98,Javanainen99,Mackie00,Hope00,Holland,Olsen,Goral00}.
 We have recently calculated that in certain parameter regimes the GPE
can give incorrect results for this process~\cite{Hope01}.  This paper
investigates the process with a three-dimensional model that includes
background gas losses, two-body losses and the atom-molecule
interactions.

The success of the GPE~\cite{GPE} might not seem so surprising
considering the effectiveness of the semiclassical approximation in
quantum optics~\cite{Walls}.  It includes the effects of s-wave
interactions, and can be readily generalised to include
multi-component condensates with inter-species couplings
\cite{Heinzen00,atomlaser}.  As a semiclassical, mean-field theory it
necessarily cannot give information about the quantum statistics of
the condensates, but for most experiments with BEC these properties
have not been observed.  Quantum statistics affect some nonlinear
quantum optical systems, the simplest of which is second harmonic
generation, where pairs of photons are coupled to single, high-energy
photons~\cite{revive}.  The analogous process in atom optics is that
of the coupling of a MBEC and a BEC, which may be done either through
tuning of a Feshbach resonance \cite{Holland}, or through
photoassociation via a two-photon Raman coupling
\cite{Heinzen00,Wynar00,Julienne98}.

The Bose-enhancement of the photoassociation of atoms from a trapped
BEC leads to giant, collective oscillations between the atomic and
molecular populations.  This enhancement of a chemical process was
dubbed ``superchemistry'' by Heinzen {\it et al.} when they first
modeled it using a two-component Gross-Pitaevskii equation (GPE)
\cite{Heinzen00}.  A more recent model using the
Hartree-Fock-Bogoliubov method includes pair correlations in the
atomic field, and showed discrepancies with the results obtained with
the GPE \cite{Holland}.  The full effects of the quantum nature of the
fields can be modeled by a set of stochastic equations for the atomic
and molecular fields based on the functional positive-P representation
\cite{Drummond80,Steel98,Drummond99,Poulsen00}.  A recent paper used
this technique in a one-dimensional calculation to show that it was
possible to see the effects of the quantum statistics in the
population dynamics of coupled atomic and molecular BEC \cite{Hope01}. 
The signature was a reduction in the transfer from atoms to molecules. 
This paper uses a more complete model which includes all three
dimensions as well as atom-molecule interactions and losses.  The
parameter regimes in which the GPE breaks down are determined.

In Sec.~\ref{sec:model} the extended model is described and the
stochastic equations of motion for the system are derived.  The
following section is a brief examination of the behaviour of the GPE and
the likely parameter regimes in which there are measurable deviations
from this behaviour.  In Sec.~\ref{sec:QS} the evolution of the atomic
population is compared with the solution of the GPE for a range of
densities and coupling strengths.

\section{Model} \label{sec:model}

\begin{figure}
\begin{center}
\epsfxsize=8cm
\epsfbox{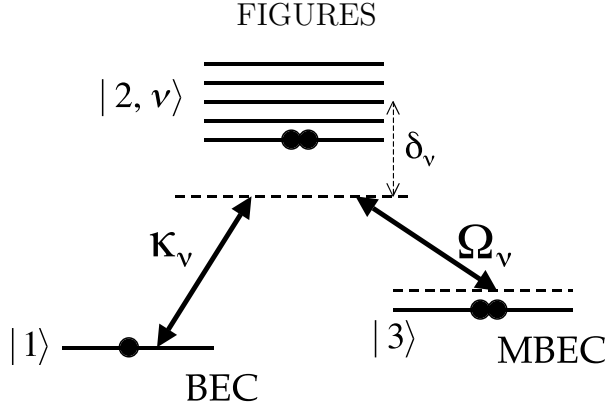}
\end{center}
\caption{Energy level scheme for coherent free-bound-bound
photoassociation.  Levels $|1\rangle$, $|2,\nu\rangle$ and $|3\rangle$
are the electronic states for the atomic BEC, the $\nu$-vibrational
level excited MBEC and the stable MBEC respectively.}
\label{fig:levels}
\end{figure}

An atomic field is coupled to a molecular field by two-colour Raman
photoassociation.  A first laser couples a single electronic level of
the atomic field to a set of molecular excited states.  A second laser
then couples these states to a stable molecular level.  The modes are
arranged as shown in Fig.~\ref{fig:levels}, with state $|1\rangle$
being the atomic BEC, state $|2,\nu\rangle$ the $\nu$th vibrational
level of the excited state of the MBEC and state $|3\rangle$ the
stable MBEC. Two laser fields induce a free-bound coupling between
$|1\rangle$ and $|2,\nu\rangle$ and a bound-bound coupling between
$|2,\nu\rangle$ and $|3\rangle$.  In a rotating frame, the Hamiltonian may
be written as
\begin{eqnarray}
    \hat{H} & = & \sum_{i=1}^{3}\left(\hat{T}_{i}+\hat{V}_{i}\right)
    +\sum_{ij}\int d^{3}x\;\hat{\psi}_{i}^{\dag}(x)
    \hat{\psi}_{j}^{\dag}(x)
    \frac{U_{ij}}{2} \hat{\psi}_{i}(x) \hat{\psi}_{j}(x) \nonumber \\
     & + & \frac{i \hbar}{2} \sum_{\nu}\int
    d^{3}x\;\left(\kappa_{\nu}(x) \hat{\psi}_{1}^{\dag\;2}(x)
    \hat{\psi}_{2,\nu}(x)-\kappa_{\nu}^{*}(x) \hat{\psi}_{1}^{2}(x)
    \hat{\psi}_{2,\nu}^{\dag}(x) \right) \nonumber \\
    & + & i \hbar \sum_{\nu} \int d^{3}x\;\left(\Omega_{\nu}(x)
    \hat{\psi}_{2,\nu}^{\dag}(x) \hat{\psi}_{3}(x) - \Omega_{\nu}^{*}(x)
    \hat{\psi}_{2,\nu}(x) \hat{\psi}_{3}^{\dag}(x)\right) \nonumber
\end{eqnarray}
where $\hat{\psi_{j}}(x)$ is the field annihilation operator for the
atomic or molecular field in state $|j\rangle$, $\hat{T}_{i}$ and
$\hat{V}_{i}$ are the kinetic and potential energy operators for the
$i^{\mbox{th}}$ field, $U_{ij}$ is the strength of the interatomic
interactions between particles in states $|i\rangle$ and $|j\rangle$,
$\kappa_{\nu}(x)$ is the Rabi frequency of the free-bound
photoassociation from level $|1\rangle$ to level $|2,\nu\rangle$ and
$\Omega(x)$ is the Rabi frequency of the bound-bound transition from
level $|2,\nu\rangle$ to level $|3\rangle$.  In this notation, the
detunings of the lasers from the bare atomic and molecular energy
levels are included in the potential energy terms $V_{2,\nu}$ and $V_{3}$.

In addition to the coherent effects produced by this Hamiltonian, the
losses from the trapped levels are included by adding standard loss
terms to the master equation.  The total master equation is:

\begin{eqnarray}
    \dot{\rho} &=& 
    -\frac{i}{\hbar}[\hat{H},\rho]+\sum_{j}\gamma_{j}^{(1)}\int 
    d^{3}x\;{\cal D}[\hat{\psi}_{j}(x)]\rho \nonumber \\
    &&+ \sum_{j} \gamma_{j}^{(2)} \int d^{3}x\; {\cal 
    D}[\hat{\psi}_{j}^{2}(x)] \rho \nonumber \\
    &&+ 2\gamma_{13}^{(2)} \int d^{3}x\; {\cal 
    D}[\hat{\psi}_{1}(x)\hat{\psi}_{3}(x)] \rho
    \label{eq:master}
\end{eqnarray}
where $\gamma_{j}^{(1)}$ is the loss rate from level $|j\rangle$ due
to background gases, $\gamma_{j}^{(2)}$ is the two-body loss rate from
level $|j\rangle$, $\gamma_{13}^{(2)}$ is the two-body loss rate due
to inelastic collisions between the BEC and MBEC fields, and where the
superoperator ${\cal D}$ is defined by
\begin{mathletters}
    \begin{eqnarray}
    {\cal D}[c] &=& {\cal J}[c] - {\cal A}[c], \\
    {\cal J}[c] \rho &=& c \rho c^{\dag},\\
    {\cal A}[c] \rho &=& \frac{1}{2}(c^{\dag}c \rho + \rho c^{\dag}c).
    \label{eq:SuperOps}
    \end{eqnarray}
\end{mathletters}

\subsection{Derivation of the stochastic field equations}

The master equation is a field operator equation with a non-trivial
level of excitation, and is therefore impossible to solve numerically
by direct means.  Analytical results are precluded by the
nonlinearities in the equations.  To find a numerical method which is
tractable in some parameter regimes, this master equation is written
in the functional positive-$P$ representation \cite{Steel98,Graham70}
\begin{equation}
       P(\{\psi^{\alpha},\psi^{\beta}\},\tau) =
       \rho^{(a)}(\{\hat{\psi},\hat{\psi}^{\dag}\},\tau)|
       _{\hat{\psi}\leftrightarrow\psi^{\alpha},
       \hat{\psi}^{\dag}\leftrightarrow\psi^{\beta}}
    \label{eq:Pplusdef}
\end{equation}
where $\rho^{(a)}$ is the density operator antinormally ordered with
respect to the field operators in the Schr\"{o}dinger picture.  It is
then possible to use the functional operator correspondences:
\begin{eqnarray}
    \hat{\psi}\rho\leftrightarrow \psi^{\alpha} P, & \rule{10mm}{0mm} &
    \hat{\psi}^{\dag}\rho \leftrightarrow \left(\psi^{\beta}-
    \frac{\partial}{\partial \psi^{\alpha}}\right) P,
    \nonumber  \\
    \rho\hat{\psi}^{\dag}\leftrightarrow \psi^{\beta} P, & \rule{10mm}{0mm} & 
    \rho\hat{\psi} \leftrightarrow \left(\psi^{\alpha}-
    \frac{\partial}{\partial \psi^{\beta}}\right) P,
    \label{eq:opcor}
\end{eqnarray}
to write a functional Fokker-Planck equation (FPE) from the master
equation.  This FPE may be written in the form:
\begin{equation}
    \frac{\partial P}{\partial t} = \sum_{\nu} \int \mbox{dx} \left[-
    \partial_{\nu} A^{\nu} + \sum_{\mu} \frac{1}{2} \partial_{\mu}
    \partial_{\nu} D^{\mu \nu}\right] P \nonumber
\end{equation}
where the elements $\mu$ and $\nu$ correspond to the $(4+2n)$
components of the fields in the positive-P representation:
$\{\psi_{1}^{\alpha}, \psi_{1}^{\beta},\ldots, \psi_{2,\nu}^{\alpha},
\psi_{2,\nu}^{\beta}, \ldots,\psi_{3}^{\alpha}, \psi_{3}^{\beta}\}$,
$A$ is the drift vector, $D$ is the diffusion matrix, and
$\partial_{(\nu\leftrightarrow
\{n,\gamma\})}\equiv\frac{\partial}{\partial \psi_{n}^{\gamma}}$.  The
drift vector is given by
\begin{eqnarray}
    A & = & \left(
    \begin{array}{c}
        \left({\cal K}_{1}-\frac{\gamma_{1}^{(1)}}{2}\right)\psi_{1}^{\alpha}
	+(-i\Gamma_{11}-\gamma_{1}^{(2)})\psi_{1}^{\beta}\psi_{1}^{\alpha 2} 
	+(-i\Gamma_{13}-\gamma_{13})\psi_{3}^{\beta}\psi_{3}^{\alpha}
	\psi_{1}^{\alpha}
	+ \sum_{\nu}\kappa_{\nu} \psi_{1}^{\beta}\psi_{2,\nu}^{\alpha}\\
	
        \left(-{\cal K}_{1}-\frac{\gamma_{1}^{(1)}}{2}\right)\psi_{1}^{\beta}
	+(i\Gamma_{11}-\gamma_{1}^{(2)})\psi_{1}^{\alpha}\psi_{1}^{\beta 2} 
	+(i\Gamma_{13}-\gamma_{13})\psi_{3}^{\alpha}\psi_{3}^{\beta}
	\psi_{1}^{\beta}
	+ \sum_{\nu}\kappa_{\nu}^{*} \psi_{1}^{\alpha}\psi_{2,\nu}^{\beta}  \\
	
	    \vdots	\\
	
	\left({\cal K}_{2,\nu}-\frac{\gamma_{2,\nu}^{(1)}}{2}
	-i\Gamma_{1;2,\nu} \psi_{1}^{\beta}\psi_{1}^{\alpha}
	-i\Gamma_{3;2,\nu} \psi_{3}^{\beta}\psi_{3}^{\alpha}
	\right) \psi_{2,\nu}^{\alpha}
	- \frac{\kappa_{\nu}^{*}}{2} \psi_{1}^{\alpha 2}
	+\Omega_{\nu} \psi_{3}^{\alpha}\\
	
        \left(-{\cal K}_{2,\nu}-\frac{\gamma_{2,\nu}^{(1)}}{2}
	+i\Gamma_{1;2,\nu} \psi_{1}^{\beta}\psi_{1}^{\alpha}
	+i\Gamma_{3;2,\nu} \psi_{3}^{\beta}\psi_{3}^{\alpha}
	\right) \psi_{2,\nu}^{\beta}
	- \frac{\kappa_{\nu}}{2} \psi_{1}^{\beta 2}
	+\Omega_{\nu}^{*} \psi_{3}^{\beta}  \\
	
	    \vdots	\\
	
        \left({\cal K}_{3}-\frac{\gamma_{3}^{(1)}}{2}\right)\psi_{3}^{\alpha}
	+(-i\Gamma_{33}-\gamma_{3}^{(2)})\psi_{3}^{\beta}\psi_{3}^{\alpha 2} 
	+(-i\Gamma_{13}-\gamma_{13})\psi_{1}^{\beta}\psi_{1}^{\alpha}
	\psi_{3}^{\alpha} - \sum_{\nu} \Omega_{\nu}^{*} \psi_{2,\nu}^{\alpha}  \\
	
        \left(-{\cal 
        K}_{3}-\frac{\gamma_{3}^{(1)}}{2}\right)\psi_{3}^{\beta}
	+(i\Gamma_{33}-\gamma_{3}^{(2)})\psi_{3}^{\alpha}\psi_{3}^{\beta 2} 
	+(i\Gamma_{13}-\gamma_{13})\psi_{1}^{\beta}\psi_{1}^{\alpha}
	\psi_{3}^{\beta} - \sum_{\nu} \Omega_{\nu}^{*} \psi_{2,\nu}^{\beta}
    \end{array}
    \right)
    \label{eq:A}  
\end{eqnarray}
where ${\cal K}_{j}=-i/\hbar (\hat{T}_{j}+\hat{V}_{j})$, and
$\Gamma_{ij}=U_{ij}/\hbar$.  The terms proportional to the density of
the excited molecular states have been dropped, as these upper states
are going to be adiabatically eliminated.  The diffusion matrix $D$ is
given by:
\begin{eqnarray}
    D & = & \left(
    \begin{array}{llcll}
            -g_{11}\psi_{1}^{\alpha 2}
            +\sum_{\nu}\kappa_{\nu} \psi_{2,\nu}^{\alpha}
        &0&\ldots&
	-g_{13}\psi_{3}^{\alpha}\psi_{1}^{\alpha}&0\\
	
        0&-g_{11}^{*}\psi_{1}^{\beta 2} 
	+ \sum_{\nu}\kappa_{\nu}^{*} \psi_{2,\nu}^{\beta}&
	\ldots&
	0&-g_{13}^{*}\psi_{3}^{\beta}\psi_{1}^{\beta}\\
	
	 \vdots&\vdots&  \ddots	&\vdots&\vdots\\
	
	-g_{13}\psi_{3}^{\alpha}\psi_{1}^{\alpha}&0
	&\ldots&
	-g_{33}\psi_{3}^{\alpha 2}&
	0\\
	
         0&-g_{13}^{*}\psi_{3}^{\beta}\psi_{1}^{\beta}&
	\ldots&
	0&-g_{33}^{*}\psi_{3}^{\beta 2}
    \end{array}
    \right)
    \label{eq:D}  
\end{eqnarray}
where $g_{ij}=\gamma_{ij}^{(2)}+i\Gamma_{ij}$ combines the atomic
interactions and the two-body losses.  The noise terms due to the self
interaction of the excited molecular states have been ignored, as the
density of these states is about to be assumed to be extremely small. 
This will enable them to be adiabatically eliminated.

The Fokker-Planck equation leads to the following set of equations for
the excited molecular states:

\begin{eqnarray}
    \frac{\partial \psi_{2,\nu}^{\alpha}}{\partial t} & = &
	\left({\cal K}_{2,\nu}-\frac{\gamma_{2,\nu}^{(1)}}{2}
	-i\Gamma_{1;2,\nu} \psi_{1}^{\beta}\psi_{1}^{\alpha}
	-i\Gamma_{3;2,\nu} \psi_{3}^{\beta}\psi_{3}^{\alpha} \right)
	\psi_{2,\nu}^{\alpha} - \frac{\kappa_{\nu}^{*}}{2}
	\psi_{1}^{\alpha 2} +\Omega_{\nu} \psi_{3}^{\alpha}\\
    \frac{\partial \psi_{2,\nu}^{\beta}}{\partial t} & = &
	\left(-{\cal K}_{2,\nu}-\frac{\gamma_{2,\nu}^{(1)}}{2}
	+i\Gamma_{1;2,\nu} \psi_{1}^{\beta}\psi_{1}^{\alpha}
	+i\Gamma_{3;2,\nu} \psi_{3}^{\beta}\psi_{3}^{\alpha} \right)
	\psi_{2,\nu}^{\beta} - \frac{\kappa_{\nu}}{2} \psi_{1}^{\beta
	2} +\Omega_{\nu}^{*} \psi_{3}^{\beta}
\end{eqnarray}

In order for two lasers to provide a two-photon transition without
populating the excited MBEC, the single photon detunings
$\delta_{\nu}$ must be made very large.  This means that the
population of the excited level is very small and that its time
derivative can be ignored.  If the terms
$\partial_{t}\psi_{2,\nu}^{\alpha,\beta}$ and
$\hat{T}\psi_{2,\nu}^{\alpha,\beta}$ in the equations of motion for
$\psi_{2,\nu}^{\alpha,\beta}$ are ignored on the basis that they are
smaller than the other terms, it becomes possible to write solutions
for $\psi_{2,\nu}^{\alpha,\beta}(x,t)$ explicitly in terms of
$\psi_{1}^{\alpha,\beta}(x,t)$ and $\psi_{3}^{\alpha,\beta}(x,t)$:
\begin{eqnarray}
    0 & = & \left(-\frac{i}{\hbar}V_{2,\nu} -\gamma_{2,\nu}^{(1)}/2
    -i\Gamma_{1;2,\nu}\psi_{1}^{\beta}\psi_{1}^{\alpha}
    -i\Gamma_{3;2,\nu} \psi_{3}^{\beta}\psi_{3}^{\alpha}\right)
    \psi_{2,\nu}^{\alpha}- \frac{\kappa_{\nu}^{*}}{2}
	\psi_{1}^{\alpha 2} +\Omega_{\nu} \psi_{3}^{\alpha}
    \label{eq:psitwo}  \\
    0 & = & \left(\frac{i}{\hbar}V_{2,\nu} -\gamma_{2,\nu}^{(1)}/2
    +i\Gamma_{1;2,\nu}\psi_{1}^{\beta}\psi_{1}^{\alpha}
    +i\Gamma_{3;2,\nu} \psi_{3}^{\beta}\psi_{3}^{\alpha}\right)
    \psi_{2,\nu}^{\beta}- \frac{\kappa_{\nu}}{2}
	\psi_{1}^{\beta 2} +\Omega_{\nu}^{*} \psi_{3}^{\beta}.
    \nonumber
\end{eqnarray}

A rearrangement of these equations, and the further assumption that
the single-photon detuning $\delta_{\nu}$ is larger than the
interaction terms $\Gamma_{i;2,\nu}
\psi_{i}^{\beta}\psi_{i}^{\alpha}$, the excited state loss rate
$\gamma_{2,\nu}^{(1)}$ and the trapping potential (i.e.
$V_{2,\nu}\approx\hbar\delta_{\nu}$), gives the following result:
\begin{eqnarray}
    \psi_{2,\nu}^{\alpha} & = & 
    \frac{i/2 \kappa_{\nu} \psi_{1}^{\alpha\;2}-i \Omega_{\nu} 
    \psi_{3}^{\alpha}}{\delta_{\nu}}
    \label{eq:psitwoexplicit}  \\
    \psi_{2,\nu}^{\beta} & = & 
    \frac{-i/2 \kappa_{\nu}^{*} \psi_{1}^{\beta\;2}+i \Omega_{\nu}^{*} 
    \psi_{3}^{\beta}}{\delta_{\nu}}.
    \nonumber
\end{eqnarray}

These equations can then be substituted into Eq.(\ref{eq:A}) and
Eq.(\ref{eq:D}) to produce the drift and diffusion matrices for the
two levels $|1\rangle$ and $|3\rangle$.  It is then possible to find a
matrix $B$ such that $D=B.B^{T}$.  In general this matrix will not be
unique.  Finally, there is a theorem which allows us to map the
solution of the master equation in Eq.(\ref{eq:master}) to the
following set of It\^{o} stochastic field equations:
\begin{eqnarray}
    \frac{\partial \psi_{1}^{\alpha}}{\partial t} & = &
        \left({\cal K}_{1}-\frac{\gamma_{1}^{(1)}}{2}
	-(g_{1}-i\Gamma)\psi_{1}^{\beta}\psi_{1}^{\alpha}
	-g_{13}\psi_{3}^{\beta}\psi_{3}^{\alpha}\right)\psi_{1}^{\alpha}
	-i\chi \psi_{1}^{\beta}\psi_{3}^{\alpha} \nonumber\\&&
        + \sqrt{-i\chi\psi_{3}^{\alpha}-(g_{1}-i\Gamma)\psi_{1}^{\alpha 2}}
	\eta_{1} - \sqrt{g_{13}\psi_{1}^{\alpha}\psi_{3}^{\alpha}}\zeta_{1}^{*}
    \nonumber  \\
     \frac{\partial \psi_{1}^{\beta}}{\partial t} & = &
        \left(-{\cal K}_{1}-\frac{\gamma_{1}^{(1)}}{2}
	-(g_{1}^{*}+i\Gamma)\psi_{1}^{\beta}\psi_{1}^{\alpha}
	-g_{13}^{*}\psi_{3}^{\beta}\psi_{3}^{\alpha}\right)\psi_{1}^{\beta}
	+i\chi^{*} \psi_{1}^{\alpha}\psi_{3}^{\beta}\nonumber\\&&
        + \sqrt{i\chi^{*}\psi_{3}^{\beta}-(g_{1}^{*}+i\Gamma)\psi_{1}^{\beta 2}}
	\eta_{2} - \sqrt{g_{13}\psi_{1}^{\beta}\psi_{3}^{\beta}}\zeta_{2}
	\label{eq:EOM3D} \\
    \frac{\partial \psi_{3}^{\alpha}}{\partial t} & = &
        \left({\cal K}_{3}-\frac{\gamma_{3}^{(1)}}{2}+i\lambda
	-g_{3}\psi_{3}^{\beta}\psi_{3}^{\alpha}
	-g_{13}\psi_{1}^{\beta}\psi_{1}^{\alpha}\right)\psi_{3}^{\alpha}
	-\frac{i}{2} \chi^{*} \psi_{1}^{\alpha 2}\nonumber\\&&
        + \sqrt{-g_{3}}\psi_{3}^{\alpha} \eta_{3} 
	+ \sqrt{g_{13}\psi_{1}^{\alpha}\psi_{3}^{\alpha}}\zeta_{1}
    \nonumber  \\
    \frac{\partial \psi_{3}^{\beta}}{\partial t} & = &
        \left(-{\cal K}_{3}-\frac{\gamma_{3}^{(1)}}{2}-i\lambda
	-g_{3}^{*}\psi_{3}^{\beta}\psi_{3}^{\alpha}
	-g_{13}^{*}\psi_{1}^{\beta}\psi_{1}^{\alpha}\right)\psi_{3}^{\beta}
	+\frac{i}{2} \chi \psi_{1}^{\beta 2}\nonumber\\ \nonumber&&
        + \sqrt{-g_{3}^{*}}\psi_{3}^{\beta} \eta_{4} 
	+ \sqrt{g_{13}^{*}\psi_{1}^{\beta}\psi_{3}^{\beta}}\zeta_{2}^{*}
\end{eqnarray}
where 
\begin{mathletters}
  \begin{eqnarray}
    \Gamma & = & \sum_{\nu}\frac{\left|\kappa_{\nu}\right|^{2}}{2 
    \delta_{\nu}},
    \label{eq:gammadef}  \\
    \chi & = & \sum_{\nu}\frac{\kappa_{\nu}\Omega_{\nu}}{\delta_{\nu}},
    \label{eq:chidef}  \\
    \lambda & = & 
    \sum{\nu}\frac{\left|\Omega_{\nu}\right|^{2}}{\delta_{\nu}},
    \label{eq:omegadef}
  \end{eqnarray}
\end{mathletters}
and the $\eta_{\nu}$ are a set of real, Gaussian noise sources which
are $\delta$-correlated in time and space:
\begin{eqnarray*}
    \overline{\eta_{i}({\bf x},t)\eta_{j}({\bf x'},t')}
    =\delta_{ij}\delta({\bf x}-{\bf x'})\delta(t-t'),
\end{eqnarray*}
and $\zeta_{\nu}$ are a set of complex, Gaussian noise sources 
which are $\delta$-correlated in time and space:
\begin{eqnarray*}
    \overline{\zeta_{i}^{*}({\bf x},t)\zeta_{j}({\bf x'},t')}
    =\delta_{ij}\delta({\bf x}-{\bf x'})\delta(t-t').
\end{eqnarray*}

These stochastic equations allow us to generate any multitime
time-normally ordered quantum field averages by averaging selected
moments of these fields over a sufficiently large sample of
trajectories\cite{Steel98}.
\begin{eqnarray}
    &\langle 
    \overleftarrow{T}\hat{\psi}^{\dag}(x,t)\ldots\hat{\psi}^{\dag}(x',t')
    \overrightarrow{T}\hat{\psi}^{\dag}(x'',t'')
    \ldots\hat{\psi}^{\dag}(x''',t''')\rangle  \nonumber\\
    &\rule{20mm}{0mm} = \overline{\psi^{\beta}(x,t)\ldots\psi_{\beta}(x',t')
        \psi^{\alpha}(x'',t'')\ldots\psi_{\alpha}(x''',t''')},\nonumber
\end{eqnarray}
where $(\overleftarrow{T})\overrightarrow{T}$ is the 
(anti-)chronological time ordering operator.

The It\^{o} stochastic equations reduce to the GPE for this system if
the noise terms are ignored.  As can be seen later, this is not the
case when using a different stochastic calculus, such as Stratonovich
calculus, which reinterprets the noise terms and therefore requires
corrections to the deterministic part of the equations.  This
reinforces the fact that the noise terms cannot be treated separately
from the deterministic terms, as the same physics produces both of
them.  It is also worth reiterating that there is no useful
interpretation of the individual trajectories of the simulation, only
the ensemble averages.

Our equations of motion are very similar to those obtainable by a
direct coupling between the BEC and the stable MBEC. The difference is
the nonlinear light shift of the BEC, which is proportional to
$\Gamma$, and the linear light shift of the MBEC, which is equal to
$\lambda$.

\section{Semiclassical superchemistry}

\begin{figure}
\begin{center}
\epsfxsize=8cm
\epsfbox{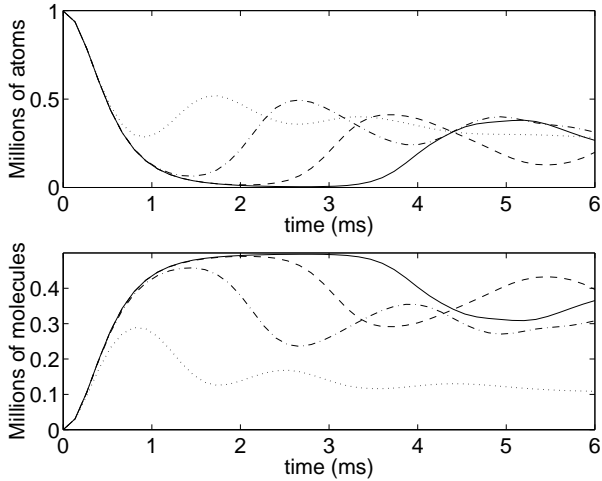}
\end{center}
\caption{Evolution of the semiclassical equations of motion for the
atomic and molecular populations in a fixed three-dimensional harmonic
trap.  The values of $\chi$ used were $3.0\times 10^{-5}$ m$^{1/2}$/s
(solid), $3.0\times 10^{-6}$ m$^{1/2}$/s (dashed), $3.0\times 10^{-7}$
m$^{1/2}$/s (dash-dotted) and $3.0\times 10^{-8}$ m$^{1/2}$/s
(dotted).}
\label{fig:semiclassical}
\end{figure}

Examination of the semiclassical equations of motion obtained by
dropping the noise terms from Eq.(\ref{eq:EOM3D}) shows that the
behaviour is sensitive to the relative strengths of the
non-linearities.  In the limit that there are no third order
nonlinearities it can be seen that if the population starts in the
atomic field then there is a complete, one-way conversion to molecules. 
In the presence of atomic and molecular interactions, the third order
nonlinearities dephase this evolution, and cause a revival of the
atomic population.  This behaviour is shown in
Fig.~\ref{fig:semiclassical}, where the evolution is shown for
different values of the atomic-molecular coupling rate.

This semiclassical behaviour has been observed in traveling-wave
second harmonic generation, which is a zero-dimensional analogue of
coupled atomic and molecular condensates.  However, it has also been
shown that the semiclassical approximation gives incorrect predictions
for the mean behaviour of the fields in traveling-wave second
harmonic generation~\cite{revive}, as well as for their quantum
statistical properties~\cite{correlations,QND}.  This discrepancy is
most pronounced when there is nearly complete conversion to the second
harmonic, which occurs when the third order nonlinearities are very
small \cite{ki3}.  It has since been shown that it is also possible to
observe the effects of the quantum nature of the fields in a
one-dimensional model of coupled atomic and molecular condensates
\cite{Hope01}.

The breakdown of the semiclassical approximation has a simple
explanation.  The semiclassical equations do not include the
spontaneous emission terms which allow a molecule to break into two
atoms in the absence of an atomic field.  To include these processes
in the model the full stochastic equations of motion must be
considered, which include all the physics of the quantum fields.  The
noise terms of the evolution mimic the dephasing effect of third-order
nonlinearities, and cause a revival of the atomic population.

For coupled atomic and molecular condensates in three dimensions, with
losses due to background gas collisions and two-body collisions, it
may be significantly more difficult to observe the quantum nature of
the fields.  In higher dimensions there is a larger volume of low
density fields than in low dimensions, which in combination with the
third order nonlinearities and the losses will tend to ``blur out''
the strong conversion to molecules which can be observed in the
semiclassical theory for low dimensions.  This effect can be
counteracted by ensuring that the field is as dilute as possible, which
makes the (second order in the fields) coupling between the BEC and
the MBEC relatively stronger than the (third order in the fields)
interaction terms.  In order to minimise the diffusion due to kinetic
energy and further reduce the effect of the interactions, the Raman
transition must be made as strong as possible.

Making the Raman transition as strong as possible requires maximum
laser intensities and a two-photon resonant coupling
($\Delta=\lambda$).  The practical limits on the total strength of the
coupling are the requirement that the excited molecular state must not
have significant population, and the need to avoid inducing large,
attractive atomic interactions through the non-linear light shift due
to the laser-assisted photoionisation.  Unless there is a low excited
state population, spontaneous losses from that highly energetic state
could disrupt the condensates, and the adiabatic approximation used to
eliminate the upper level may break down.  This limit imposes the
condition that the single photon detuning $\delta$ must be
significantly larger than the Rabi frequencies of the individual laser
beams.  The second requirement means that the non-linear light shift
should stay of the order of the repulsive interactions.  If the Rabi
frequency of the first laser is chosen such that $\Gamma =
\Gamma_{11}$, then the nonlinear light shift will completely cancel
the interactions between the atoms.  There will still be repulsive
interactions between the molecules and attraction between the atoms
and the molecules.

Dilute condensates can be created by evaporating below the BEC
condition using strong straps, and then adiabatically lowering the
trap strengths in all directions.  The trap depth remains large
compared to the chemical potential of the condensate even for
extremely weak traps \cite{trapdepth}.  The field cannot be
arbitrarily dilute, however, as the accurate measurement of the fields
depends on a sufficient column density.  Many current experiments use
highly anisotropic ``cigar-shaped'' traps, which allow a high column
density along the long axis.  For a given aspect ratio
$\omega_{x}=\omega_{y}=A \omega_{z}$ and number of trapped atoms $N$,
it is possible to calculate the transverse trap frequency required to
obtain the required peak column density $\sigma_{p}$:
\begin{equation}
    \omega_{x}=\left(\frac{18 \;\sigma_{p}^{5} \;U_{11}^{2}\; 
    \pi^{3}}{125 \;N^{3}\; m^{2}\; A^{2}}\right)^{\frac{1}{4}}.
    \label{eq:weakomega}
\end{equation}
The peak density scales inversely with the aspect ratio $A$, which has
been as large as $100$ in current experiments \cite{aspectratio}.  In
the next section the behaviour of the quantum model of the system will
be investigated in both dilute and strong traps, and with both weak and
strong coupling.

\section{Quantum superchemistry} \label{sec:QS}

This section will examine the difference between the quantum solution
and the semiclassical solution allowing for physical constraints such
as limited detuning and laser power.  To ensure that the adiabatic
approximation is being made self-consistently, the density of atoms in
the excited state (which must remain small) can be calculated from the
amplitudes of the other fields.

Since our last calculation \cite{Hope01}, the interaction between the
atomic and molecular species has been measured by examining the
resonances in the molecular formation due to photoionisation
\cite{Wynar00}.  The atoms are attracted to each other with a
scattering length of $a_{13}=-180\pm 150$~$a_{o}$, which leads to a
value of $U_{13}=6\,\pi\,a_{13}\,\hbar^{2}/m = -1.4\times 10^{-50}$~J~m$^{3}$.  The strength of the interactions between the Rb$_{2}$
molecules has not been measured, so it will be assumed that they are
the same as the interatomic interactions for the purposes of this
calculation ($U_{11}=U_{33}=3\times 10^{-51}$~J~m$^3$).

The losses due to background gases are very slow on the timescale of
the superchemistry.  All calculations in this work use a value of
$\gamma_{1}^{(1)}=\gamma_{3}^{(1)}=0.01 s^{-1}$, which is a
comfortable overestimate of the background gas losses measured by the
group at JILA \cite{Burt97}.  At the high densities involved in the
production of an atomic BEC, the three-body losses are dominant and
the two-body losses are negligible ($\gamma_{1}^{(2)}\le 1.6\times
10^{-22}$ m$^{3}/s$).  By contrast, the three-body losses can be
ignored at the low densities considered in this work, but in the
presence of molecules the two-body losses cannot be ignored, as there
is a significant cross-section for the inelastic scattering of an atom
from a molecule.  This was recently measured to be
$\gamma_{13}^{(2)}<8\times 10^{-17}$ m$^{3}/s$, and in this paper the
upper bound shall be used in all calculations.

It is experimentally difficult to produce large Rabi frequencies for
the atom-molecule interaction due to the low Franck-Condon factors,
but this does not appear to be a limitation of this system provided
large Rabi frequencies can be achieved for the molecule-molecule
transition.  

\subsection{Numerical methods}

Although It\^{o} equations of motion for this system have been
produced in the previous section, it is much easier to use high-order
integration methods on Stratonovich equations.  This is because the
normal chain rule of differentiation applies to Stratonovich calculus,
and standard high-order methods can be used without modification.  The
transformation from a set of It\^{o} equations to a set of
Stratonovich equations can be performed easily, as it involves a
simple correction to the deterministic terms.  Unfortunately, this
correction is infinite for Eq.(\ref{eq:EOM3D}).  This is a recurrence
of the renormalisation problem often found in quantum field theories.

The infinite correction only occurs in the continuous field
description.  When the conversion to Stratonovich calculus is done
after the fields have been discretised and placed on a rectangular
grid with spacing of $\Delta x$, $\Delta y$ and $\Delta z$ in the
three dimensions, the following correction terms are obtained:
\begin{eqnarray}
    \left.\left(
    \begin{array}{c}
        \frac{\partial \psi_{1}^{\alpha}}{\partial t}  \\
        \frac{\partial \psi_{1}^{\beta}}{\partial t}  \\
        \frac{\partial \psi_{3}^{\alpha}}{\partial t}  \\
        \frac{\partial \psi_{3}^{\beta}}{\partial t}
    \end{array}\right)\right|_{\mbox{correction}}
    & = & 
    \left(\begin{array}{c}
     \frac{g_{13}/2 + g_{1}}{2\,\Delta x\,\Delta y\,\Delta z} \psi_{1}^{\alpha}  \\
        \frac{g_{13}^{*}/2 + g_{1}^{*}}{2 \,
    \Delta x\,\Delta y\,\Delta z} \psi_{1}^{\beta}  \\
        \frac{g_{13}/2 + g_{3}}{2 \,
    \Delta x\,\Delta y\,\Delta z} \psi_{3}^{\alpha}  \\
        \frac{g_{13}^{*}/2 + g_{3}^{*}}{2 \,
    \Delta x\,\Delta y\,\Delta z} \psi_{3}^{\beta}
    \end{array}\right).
    \label{eq:Stratcorrection}
\end{eqnarray}
This correction is equivalent to a stepsize-dependent phase shift of the
fields.  It is clear that the correction is infinite in the limit of
zero spatial cell size.  In fact, this limit is not realistic for
atomic fields, as the interactions between the particles are modeled
as a contact potential, and this approximation fails at a sufficiently
small distance scale.  This puts a lower limit on the distance between
the gridpoints, which will be of the order of magnitude of the
scattering length.  This forced discretisation is equivalent to the
cutoff in momentum which is usually used to solve renormalisation
problems.

In a numerical simulation with stepsize $\Delta t$, the noise terms
$\eta({\bf x},t)$ are included at each timestep by choosing a random
number $R$ from a Gaussian distribution centred around zero and with
unit width.  The noise terms are then formed using $\eta(x,t)=R
/\sqrt{\Delta x\,\Delta y\,\Delta z\; \Delta t}$.  This means that
going to smaller grid spacing slows the computation in three ways. 
The field contains more points, which proportionally increases the
computation time per timestep.  Also, the noise terms become larger,
which means that both a smaller timestep is required to perform a
stable integration and a larger ensemble of paths is required to
obtain precise averages.  It follows that these stochastic methods are
most successful when the density is low.  Fortunately, the strong
effect should also be in this limit.

The stochastic integration was performed with the XMDS package,
developed by Collecutt and Drummond \cite{XMDS}.  In collaboration
with them and the ANU Supercomputer Facility, this has been adapted to
run multiple integrations of the stochastic equations in parallel on
the APAC supercomputer \cite{APAC}.

\subsection{Quantum superchemistry with varying coupling rates}

The feasibility of observing quantum field effects shall be
demonstrated by considering the most dilute system possible.  The
diluteness will be limited by the requirement that the column density
be large enough to provide easy measurements, and by the total number
of atoms available in the initial trap.  All further calculations
assume that it will be possible to have a million atoms in the trap
after the adiabatic expansion.  The noise in the column density
measurements is estimated to be $10^{13}$ m$^{-2}$, which is that of
the current BEC at ANU \cite{Lye01}.  For the purposes of this
subsection, it is assumed that the trap has an aspect ratio of $A=30$. 
The weakest trap in which $99.5\%$ of the atoms are above the noise
level in the column density has a trap frequency of
$\omega_{x}=0.59$s$^{-1}$, and a condensate size of $0.09$mm in the
strong axis and $0.3$mm in the weak direction.  The effect of varying
the coupling rate in such a trap will be calculated in this
subsection.  The next subsection will examine the quantum field
effects in stronger traps.

\begin{figure}
\begin{center}
\epsfxsize=8cm
\epsfbox{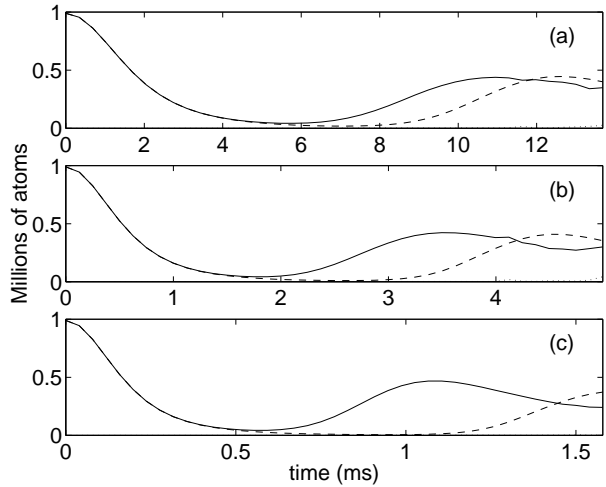}
\end{center}
\caption{Evolution of the atomic population (solid lines) compared to
the solution of the GPE (dashed lines).  Trap parameters are as given
in the text.  In each case the atomic-molecular coupling $\kappa$ was
chosen such that $\Gamma=\Gamma_{11}$, and the detuning $\delta$ was
chosen to be $50$ times larger than the molecule-molecule Rabi
frequencies $\Omega$.  Thus, a larger coupling strength requires a
higher detuning.  (a) $\chi=4.7\times 10^{-6}$m$^{3/2}$ s$^{-1}$,
which corresponds to a detuning of $160$MHz, (b) $\chi=1.5\times
10^{-5}$m$^{3/2}$ s$^{-1}$, which corresponds to a detuning of
$1.6$GHz, (c) $\chi=4.7\times 10^{-5}$m$^{3/2}$ s$^{-1}$, which
corresponds to a detuning of $16$GHz.  The dotted lines, where
visible, are the uncertainties in the theory due to the sampling
error.}
\label{fig:QSCCoupling}
\end{figure}

Fig.(\ref{fig:QSCCoupling}) compares the evolution of the atomic
population with the solution to the Gross-Pitaevskii equation.  It can
be seen that the effect of the quantum statistics of the field grows
as the coupling increases.  The coupling cannot be increased
indefinitely, due to restrictions on both the available laser power
and allowable detuning.

The difference between the dynamics of the system and the
semiclassical result is essentially unaffected by the losses, which
dominate at higher densities and over longer timescales.  By contrast,
the inclusion of the atom-molecule interactions in this model has
reduced the distinction at high densities.  From
Fig.(\ref{fig:QSCCoupling}) it can be seen that the deviation from the
semiclassical result is clearly observable for currently achievable
trap and coupling parameters.

\subsection{Quantum superchemistry at varying densities}

Although it is possible to use very weak traps to work in the low
density limit, most experiments are performed using stronger traps
than those modeled in the previous section.  This subsection will
examine the effects of changing the density.  

The most dramatic effect of increasing the strength of the trap in the
model is that the calculation becomes less stable.  As previously
explained, the noise in the numerical simulation becomes relatively
larger as the unit cell size decreases.  Stronger traps mean that the
overall size of the condensate decreases, and so does the minimum
allowable stepsize in the spatial grid.  It therefore becomes harder
to use the stochastic field equations to model the system for long
times.  The stochastic method will always produce results up to a
certain point in time, but as the individual trajectories become
unstable numerically it becomes impossible to integrate further.  

Under the constraint that there are just enough atoms to provide
easily measurable results, the ratio of the second-order and
third-order nonlinearities scales as $\omega_{x}^{2/3}$, whereas the
condensate volume will scale as $\omega_{x}^{-2}$.  The difference
between the full, quantum evolution of the system and the
semiclassical evolution depends on the ratio of the second-order and
third-order nonlinearities, so based on this scaling the distinction
may be measurable for trap strengths where the positive-P method
cannot calculate the correct result.

\begin{figure}
\begin{center}
\epsfxsize=8cm
\epsfbox{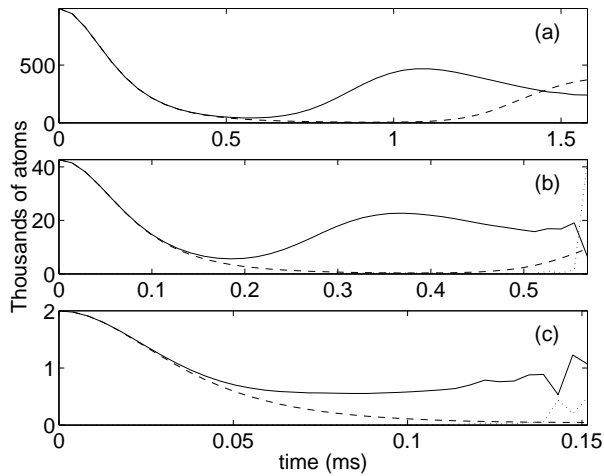}
\end{center}
\caption{Evolution of the atomic population (solid lines) compared to
the solution of the GPE (dashed lines).  The parameters used are
$\Gamma=\Gamma_{11}$, $\chi=4.7\times 10^{-5}$m$^{3/2}$ s$^{-1}$,
which corresponds to a detuning of $16$GHz.  The trap has an aspect
ratio of $A=30$, and trap frequencies (a) $\omega_{x}=2 \pi \times
0.092$Hz, (b) $\omega_{x}=2 \pi \times 1.0$Hz, and (c) $\omega_{x}=2
\pi \times 10$Hz.  The dotted lines, where visible, are the
uncertainties in the theory due to the sampling error.}
\label{fig:QSCDensity}
\end{figure}

Fig.~(\ref{fig:QSCDensity}) shows the evolution of the atomic
population for three different trap frequencies.  The aspect ratio of
the trap and the coupling rate are kept the same.  The number of
atoms in the trap is varied so that each trap contains the minimum
number of atoms required to maintain $99.5\%$ of them above a column
density of $\sigma=10^{13}$ m$^{-2}$.  An approximate method of
calculating this is that the number of atoms $N$ required to obtain a
peak density of $\sigma_{p}$ is
\begin{equation}
    N = \left(\frac{18\;U_{11}^{2}\;\sigma_{p}^{5}}
    {125\;\omega_{x}^{4}\;M^{2}\;A^{2}}\right)^{1/3}.
    \label{eq:nperomega}
\end{equation}
The figure shows a significant effect of the quantum statistics of the
field even for much higher densities.  Unfortunately, it also shows
the difficulty in performing quantitative experiments at these
densities, where the theoretical results are difficult or impossible
to obtain by the stochastic methods used in this paper.  In each plot,
the integration cannot be performed far beyond the region shown.

The importance of including the quantum statistical effects
actually appears to increase with density.  This is surprising, as we
would expect the third-order nonlinearities to be the main source of
dephasing of the conversion, and they are getting stronger with
density.  This may be due to the fact that the densities considered
here are so low that the main dephasing is being caused by the kinetic
energy term.  If this is true, then there will be an optimal density
which will maximise the parametric conversion rate compared to the
kinetic energy terms, but minimise the two-body interaction terms with
respect to the conversion rate.  Detailed testing of this hypothesis 
is difficult due to the numerical difficulties, and is beyond the 
scope of this work.

\section{Conclusions}

This paper has shown that it is possible to observe the quantum
statistical effects of the field in coupled atomic and molecular BEC.
The effect is most clearly visible in the low-density and
high-coupling strength limits.  

Although it is a very successful model, the GPE cannot be applied to
every system of coupled BECs.  While it may seem reasonable to expect
that the quantum statistics will tend to affect the multi-time
correlations of the field rather than the mean field, our result shows
it is sometimes also important to include them when considering the
equations of motion for moments of the mean field.  The signature of
the breakdown of the GPE occurs in the simplest experimental
observable - the total atomic and molecular populations.

The calculations in this paper were numerically intensive.  Although
the positive-P representation contains the full description of the
quantum field, it cannot be applied to every system of coupled BECs. 
Individually, trajectories do not have to behave in a physical fashion
and they can become unstable over time in the absence of damping.  The
technique is therefore often only useful over short timescales, or for
examining systems which are well described by the coherent state basis
which underlies the description.  This is not usually a natural basis
of the atomic field, although it is often a good basis for the optical
field.  A possible counterexample may be a continuously pumped atom
laser operating well above threshold, for which the positive-P
representation may be tractable.  It is also to be expected that the
behaviour of such a device would depend critically on the quantum
statistics.

\acknowledgments

This research was supported by the Australian Research Council.  I
would like to thank Greg Collecutt and Peter Drummond for their useful
numerical integration package XMDS, which I recommend.  I would like
to thank Greg again, as well as Margaret Kahn and David Singleton from
the ANU Supercomputer Facility for their help in adapting it to work
on a parallel computer.  I would also like to thank Craig Savage, John
Close, Jessica Lye and Cameron Fletcher for their informative
discussions.

\end{document}